\documentstyle[psfig]{mn}

\newif\ifAMStwofonts

\newcommand{\etal}{et al.\ }
\newcommand{\ie}{i.e.\ }
\newcommand{\eg}{e.g.\ }
\newcommand{\GC}{Galactic Centre}
\newcommand{\SgrAEast}{Sgr~A East}
\newcommand{\NH}{$N_{\rm H}$}
\newcommand{\NHUNIT}{H~cm$^{-2}$}
\newcommand{\FX}{$F_{\rm X}$}
\newcommand{\FLUXUNIT}{erg~s$^{-1}$~cm$^{-2}$}

\newcommand{\Einstein}{\it Einstein}
\newcommand{\ROSAT}{\it ROSAT}
\newcommand{\ASCA}{\it ASCA}
\newcommand{\SAX}{\it SAX}
\newcommand{\BSAX}{\it Beppo-SAX}
\newcommand{\XMM}{\it XMM}
\newcommand{\XMMN}{\it XMM-Newton}
\newcommand{\Chandra}{\it Chandra}
\newcommand{\XMMfil}{XMM~J174540$-$2904.5}

\newcommand{\XMMJsoft}{XMM~J174545.5$-$285830}

\ifoldfss
  \ifCUPmtlplainloaded \else
    \NewTextAlphabet{textbfit} {cmbxti10} {}
    \NewTextAlphabet{textbfss} {cmssbx10} {}
    \NewMathAlphabet{mathbfit} {cmbxti10} {} 
    \NewMathAlphabet{mathbfss} {cmssbx10} {} 
  \fi
  \ifAMStwofonts
    \ifCUPmtlplainloaded \else
      \NewSymbolFont{upmath} {eurm10}
      \NewSymbolFont{AMSa} {msam10}
      \NewMathSymbol{\upi}     {0}{upmath}{19}
      \NewMathSymbol{\umu}     {0}{upmath}{16}
      \NewMathSymbol{\upartial}{0}{upmath}{40}
      \NewMathSymbol{\leqslant}{3}{AMSa}{36}
      \NewMathSymbol{\geqslant}{3}{AMSa}{3E}

       \let\le=\leqslant
       
    \fi
  \fi
\fi 

\ifnfssone
  \newmathalphabet{\mathit}
  \addtoversion{normal}{\mathit}{cmr}{m}{it}
  \addtoversion{bold}{\mathit}{cmr}{bx}{it}
  \newmathalphabet{\mathbfit} 
  \addtoversion{normal}{\mathbfit}{cmr}{bx}{it}
  \addtoversion{bold}{\mathbfit}{cmr}{bx}{it}
  \newmathalphabet{\mathbfss} 
  \addtoversion{normal}{\mathbfss}{cmss}{bx}{n}
  \addtoversion{bold}{\mathbfss}{cmss}{bx}{n}
  \ifAMStwofonts
    \ifCUPmtlplainloaded \else
      \UseAMStwoboldmath
      \makeatletter
      \new@mathgroup\upmath@group
      \define@mathgroup\mv@normal\upmath@group{eur}{m}{n}
      \define@mathgroup\mv@bold\upmath@group{eur}{b}{n}
      \edef\UPM{\hexnumber\upmath@group}
      \new@mathgroup\amsa@group
      \define@mathgroup\mv@normal\amsa@group{msa}{m}{n}
      \define@mathgroup\mv@bold\amsa@group{msa}{m}{n}
      \edef\AMSa{\hexnumber\amsa@group}
      \makeatother
      \mathchardef\upi="0\UPM19
      \mathchardef\umu="0\UPM16
      \mathchardef\upartial="0\UPM40
      \mathchardef\leqslant="3\AMSa36
      \mathchardef\geqslant="3\AMSa3E

       \let\le=\leqslant

    \fi
  \fi
\fi 

\ifnfsstwo
  \DeclareMathAlphabet{\mathbfit}{OT1}{cmr}{bx}{it}
  \SetMathAlphabet\mathbfit{bold}{OT1}{cmr}{bx}{it}
  \DeclareMathAlphabet{\mathbfss}{OT1}{cmss}{bx}{n}
  \SetMathAlphabet\mathbfss{bold}{OT1}{cmss}{bx}{n}
  \ifAMStwofonts
    \ifCUPmtlplainloaded \else
      \DeclareSymbolFont{UPM}{U}{eur}{m}{n}
      \SetSymbolFont{UPM}{bold}{U}{eur}{b}{n}
      \DeclareSymbolFont{AMSa}{U}{msa}{m}{n}
      \DeclareMathSymbol{\upi}{0}{UPM}{"19}
      \DeclareMathSymbol{\umu}{0}{UPM}{"16}
      \DeclareMathSymbol{\upartial}{0}{UPM}{"40}
      \DeclareMathSymbol{\leqslant}{3}{AMSa}{"36}
      \DeclareMathSymbol{\geqslant}{3}{AMSa}{"3E}

       \let\le=\leqslant

    \fi
  \fi
\fi 

\ifCUPmtlplainloaded \else
  \ifAMStwofonts \else 
    \def\upi{\pi}
    \def\umu{\mu}
    \def\upartial{\partial}
  \fi
\fi

\title{XMM-Newton observations of Sagittarius A East}
\author[M. Sakano, R.S Warwick, A. Decourchelle and P. Predehl]
       {M. Sakano,$^1$\thanks{Japan Society for the Promotion of Science
       (JSPS).} R. S. Warwick$^1$, A. Decourchelle$^2$ and P. Predehl$^3$\\
        $^1$Department of Physics and Astronomy, University of Leicester,
	Leicester LE1 7RH, UK\\
	$^2$CEA/DSM/DAPNIA, Service d'Astrophysique, C.E. Saclay, 
	91191 Gif-sur-Yvette Cedex, France\\
	$^3$Max-Planck-Institut f\"ur extraterrestrische Physik,
	Postfach 1312, D-85741 Garching, Germany}
\date{Accepted 2003 ???? ??. Received 2003 ???? ??}

\pagerange{\pageref{firstpage}--\pageref{lastpage}}
\pubyear{2003}

\begin{document}

\maketitle

\label{firstpage}

\begin{abstract}
We present an analysis of a recent {\it XMM-Newton} observation of 
Sgr~A East, a supernova remnant located close to the Galactic Centre.  
Very high quality X-ray spectra reveal many emission lines from
highly ionized atoms consistent with a multi-temperature
thin thermal plasma in ionization equilibrium. We use a two-temperature 
model to fit the spectra and derive temperatures of 1~keV and 4~keV. 
There is significant concentration of iron towards the centre of the X-ray 
source  such that the iron abundance varies from $\sim$4 times solar 
in the core down to $\sim 0.5$ solar in the outer regions, which contrasts
with the rather uniform distribution of other metals such as
sulfur, argon and calcium, which have abundances in the range 1--3.
The derived total energy, mass, and the abundance pattern are consistent with a
single supernova event, either of type-Ia or type-II origin, involving
a relatively
low-mass progenitor star.  A weak 6.4-keV neutral iron fluorescence line  
is also detected, the 
illumination source most likely being {\SgrAEast} itself.  The  
morphology and spectral characteristics of {\SgrAEast} show no
clear linkage to putative past activity in Sgr~A$^*$.
\end{abstract}


\section{INTRODUCTION}

Sagittarius~A (Sgr~A) is an extremely bright radio source
situated at the {\GC}. At its core is Sgr~A$^*$, a region which
harbours a supermassive black hole with a mass of 2--3 $\times 10^6$
M$_{\sun}$ ({\eg} Genzel {\etal} 2000).  Many reviews have addressed
the variety of structures, which are apparent on a wide range of spatial 
scales, and the complex web of interactions which characterise this region
({\eg} Yusef-Zadeh {\etal} 2000; Falcke {\etal} 1999;
Mezger, Duschl \& Zylka 1996).

The Sgr~A complex consists of Sgr~A West and {\SgrAEast}.  Sgr~A West
includes Sgr~A$^*$, a three-arm spiral-like structure (the mini-spiral)
in orbit around Sgr~A$^*$, and the central star cluster (IRS~16).
On the plane of the sky, the {\SgrAEast} radio source encompasses Sgr~A West 
and has a non-thermal shell-like structure \cite{Ekers1975}.  This morphology
has been explained in terms of a supernova remnant (SNR) 
({\ie}SNR G0.0+0.0; Jones 1974; Ekers {\etal} 
1983; Green 2001).  However, alternative and more exotic interpretations
have also been proposed, {\eg} this is the remnant of an outflow
triggered by an explosion in Sgr~A$^*$.  It is certainly the case that if 
{\SgrAEast} has a direct physical link to Sgr~A West and Sgr~A$^*$, then its 
study is particularly interesting in the context of past activity of the 
central supermassive  black hole. But equally well, if {\SgrAEast} is simply 
an SNR, then its study should lead us to a better understanding of the special 
interstellar environment of the {\GC} region.

The non-thermal shell of {\SgrAEast} is elongated nearly parallel to the
Galactic plane with an overall size scale of 3{\farcm}5$\times$2{\farcm}5 
(8$\times$6 pc$^2$ for a 8.0~kpc distance).  It is surrounded by a dust ring
\cite{Mezger1989} and, in projection, overlaps with the giant molecular clouds
M$-$0.02$-$0.07 (the 50~km~s$^{-1}$ molecular cloud; {\eg} Serabyn, Lacy
\& Achtermann 1992) and M$-$0.13$-$0.08 (the `20~km~s$^{-1}$' cloud;
{\eg} Mezger {\etal} 1986).  The morphology of the dust
ring and the cloud M$-$0.02$-$0.07, as well as the detection of 
coincident OH maser emission (Yusef-Zadeh {\etal} 1996, 1999), strongly 
suggests that the non-thermal shell physically interacts with the dust ring 
and cloud. Based on this observational result, Yusef-Zadeh \& Morris (1987) 
proposed a model in which a supernova occurred inside the molecular cloud and
created the {\SgrAEast} shell.
More recently Yusef-Zadeh {\etal} (2000) have outlined a picture in which
Sgr~A West is embedded in the frontmost region of {\SgrAEast} and both 
lying within an ionized gas halo of dimension $\sim$4~arcmin ($\sim$9~pc)
(Anantharamaiah, Pedlar \& Goss 1999).

Optical, ultraviolet, and soft X-ray observations of
{\SgrAEast} are curtailed due to heavy interstellar absorption in the line
of sight, whereas hard X-ray to $\gamma$-ray observations have provided only
limited information due to the lack of instruments affording sufficient 
spatial resolution.  Mayer-Hasselwander {\etal}
\shortcite{Mayer1998} detected a significant GeV $\gamma$-ray source
towards the {\GC} with EGRET with a spatial accuracy of 0.2 deg (2EG J1746$-$2852 =
3EG J1746$-$2852, re-designated in Hartman {\etal} 1999).  Melia {\etal}
(1998a,~b) and Fatuzzo \& Melia (2003) suggest that the $\gamma$-rays may originate in {\SgrAEast}, 
although other possibilities,  for example the candidate source may 
be Sgr~A$^*$ or other point/diffuse sources, cannot be excluded.  
In the soft X-ray band below 3~keV, although major part of the emission
is absorbed, {\Einstein} \cite{Watson1981} and
{\ROSAT} (Predehl \& Tr\"{u}mper 1994) have detected a diffuse emission 
extending for $\sim$20 arcmin.  The detailed structure in the close
vicinity of Sgr~A was left unknown.
In the harder X-ray band of 2--10 keV, with the moderate spatial resolution of 
{\ASCA} (the half power diameter (HPD) of 3 arcmin), Koyama {\etal} 
\shortcite{Koyama1996} were not able to resolve 
Sgr~A$^*$, but did see extended emission ($2'\times3'$) with a peak in surface brightness 
coinciding with Sgr~A$^*$.  The spectrum of the extended emission was found 
to have many emission lines from highly ionized atoms including both
helium(He)-like and hydrogen(H)-like iron.  The overall spectral
form is very similar to that of the hot plasma which is observed over the 
entire {\GC} region except for the absence of a 6.4-keV line from neutral 
iron (Koyama {\etal} 1996; Maeda 1998; Sakano {\etal} 2002).  A {\BSAX} 
observation also confirmed the {\ASCA} result \cite{Sidoli_M1999}.

More recently this region has been observed by {\Chandra}/ACIS in the
hard X-ray band at
a spatial resolution of 0.5 arcsec (Baganoff {\etal} 2003; Maeda {\etal}
2002) and  Sgr~A$^*$ was resolved in the X-ray band for the first time
\cite{Baganoff2003}.   Intriguingly Baganoff 
{\etal} (2004) have claimed on the basis of the most recent 0.5~Ms 
{\Chandra} observation,  that during quiescence the X-ray emission detected
from Sgr~A$^*$ is slightly extended. However,
the peculiar X-ray flares seen 
from the exact location of Sgr~A$^*$ in both {\Chandra} (Baganoff
{\etal} 2001) and {\XMMN} observations (Goldwurm {\etal} 2002; Porquet {\etal} 2003), are almost 
certainly modest luminosity outbursts on Sgr~A$^*$ itself. 

In the {\Chandra} observation, bright diffuse X-ray emission from
{\SgrAEast} was also detected and resolved \cite{Maeda2002}.
The X-ray emission associated with {\SgrAEast} fills the inner part
of the non-thermal radio shell and has a spectrum 
characteristic of thin thermal emission at a temperature 
$kT\sim 2$ keV and a metal abundance $Z\sim 4$ \cite{Maeda2002}.

As one of the Guaranteed Time programmes, we have conducted a survey
of the {\GC} region with {\XMMN}, including the field of Sgr~A$^*$.  
The {\XMMN} mirrors and CCD imaging instruments offer an excellent effective 
area as well as good energy resolution (${\Delta}E/E\sim0.02$ at 5.9~keV) and 
moderate spatial resolution ($\sim$5~arcsec).  With this capability, 
we obtained spectra of excellent quality from {\SgrAEast}.  The first 
quick-look result of {\SgrAEast} and its surroundings has been reported in 
Sakano {\etal} (2003b).  In this paper, we report our detailed results  
on {\SgrAEast} as derived from the {\XMMN} observations and discuss their 
implications.  We adopt a distance of 8.0~kpc to the {\GC}
throughout this paper \cite{Reid1993}.

\section[]{THE OBSERVATIONS}

We focus on the {\XMMN} observation of the Sgr A region carried out on
2001 September 4 as part of a survey of the whole {\GC} region
(see Warwick 2002 and Sakano {\etal} 2003c) and specifically concentrate on
the results from the European Photon Imaging Camera (EPIC).  The EPIC 
instrument comprises of two MOS CCD cameras (MOS1 and 2;
Turner {\etal} 2001) and one pn CCD camera \cite{Struder2001}, which work
simultaneously.  The MOS and pn CCDs were operated in Full Frame
and Extended Full Frame mode, respectively, with the medium filter 
selected.

We have carried out the data reduction and filtering with the Standard
Analysis Software ({\sc sas}) Ver.5.4.  A preliminary screening of
the data was applied to exclude intervals of high instrumental background
(based on the full-field light curve above 10~keV, which is dominated by 
particle background events).  The effective exposure times for MOS1, 2
and pn after filtering are 23.0~ks, 23.1~ks, and 19.7~ks, respectively.

For each event, we accepted the pixel patterns of 0--12 (single to
quadruple events) for MOS.  For pn, we only accepted the single event
(pattern of 0) because of the calibration uncertainties for other pixel
patterns in the pn Extended Full Frame mode.

\section[]{RESULTS}

\subsection[]{Image Analysis\label{sec:image}}

\begin{figure*}
 \psfig{file=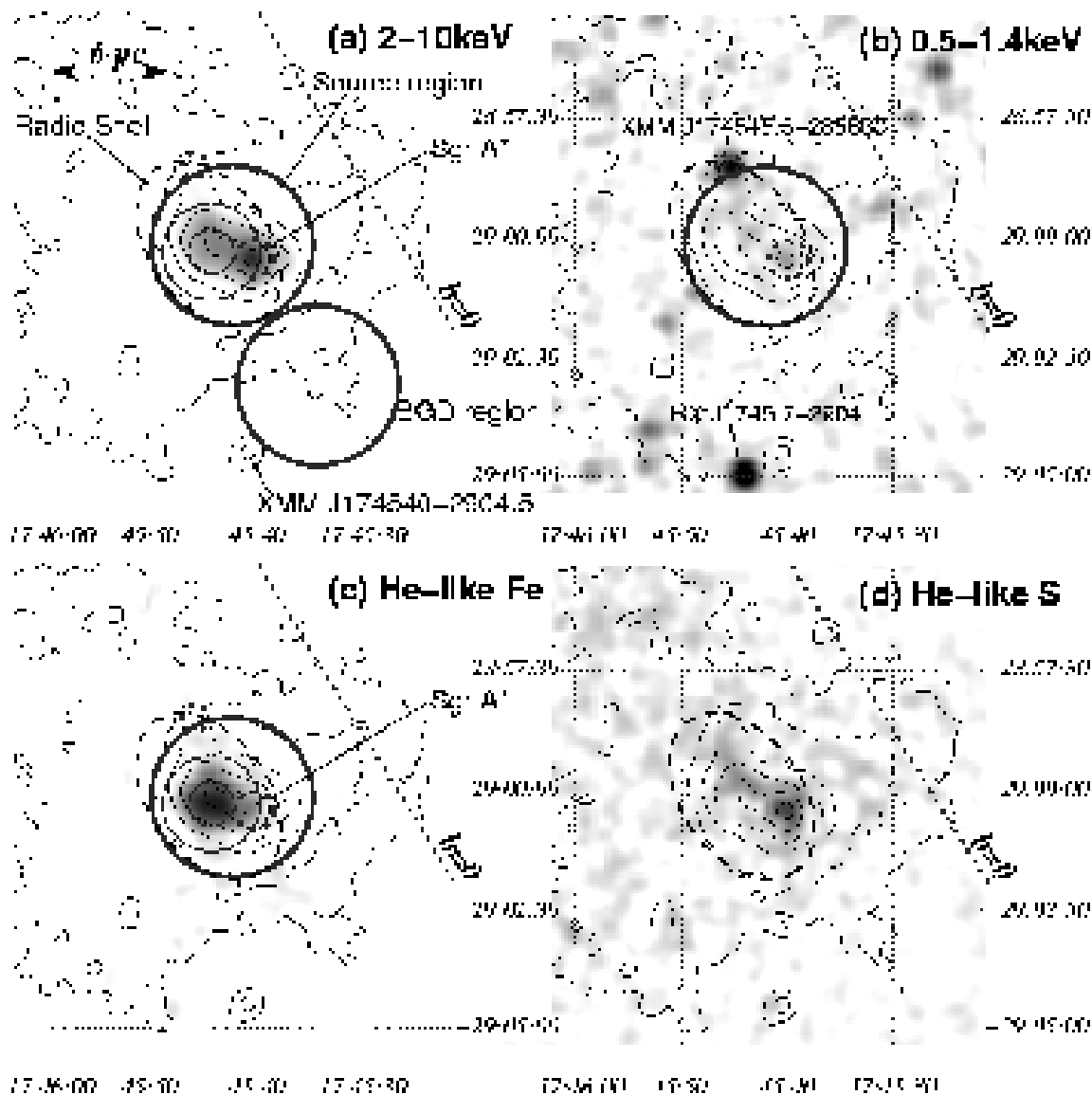,width=0.98\textwidth,clip=}
 \caption{Greyscale images of {\SgrAEast} and its surroundings measured
 with the MOS 1+2 cameras in various energy ranges as follows: (a)
 2--10 keV band; (b) 0.5--1.4 keV; (c) a narrow bandpass centred on
 the 6.7 keV He-like iron line; (d) a narrow bandpass centred on the 2.4 keV
 He-like sulphur line.  Each image is overlaid with the J2000 
 coordinates (thin dotted lines), the Galactic Plane ($b_{\rm II}=0^{\circ}$) 
 (thick dotted line) and the position of the oval radio shell (thin dashed 
 line).  The contours trace the 2--10 keV band surface brightness.
 The data have been smoothed with a Gaussian filter with $\sigma=$4
 arcsec.  In the line images (c, d), the underlying continuum has been 
 estimated from an adjacent energy band image (assuming an averaged spectral 
 shape -- see text) and subtracted.  The two large circles (thick solid lines) 
 with radii of 100~arcsec show the regions used to accumulate the primary 
 source and background spectra, but excluding the contribution of two discrete
 sources, namely Sgr~A$^*$ and  {\XMMJsoft} identified by name (and by the 
 two small circles drawn with thick dotted lines) in panels (a) \& (b).  
 The two circles (thin solid lines) with radii of 28~arcsec and 60~arcsec 
 (a, c) correspond to sub-regions of {\SgrAEast} discussed in the text.  
 The positions of a hard X-ray filament ({\XMMfil}; Sakano {\etal}
 2003a) and a soft X-ray source RX J1745.7$-$2904 (identified as a
 star GSC 06840$-$00590) are also marked.
 \label{fig:img}}
\end{figure*}

Fig.~\ref{fig:img}a and b show images of the region encompassing the Sgr~A 
complex measured, respectively, in the hard (2--10 keV) and soft 
(0.5--1.4 keV) X-ray bands with the MOS1+2 cameras.
The highest surface brightness feature in the hard band coincides with
the location of Sgr~A$^*$, the central massive black hole.  We find that 
the emission around Sgr~A$^*$ is not point-like and that Sgr~A$^*$ itself
is not resolved.  Goldwurm et al. (2002) have reported the occurrence
of an X-ray flare from Sgr~A$^*$ at the very end of this observation, but 
here we have specifically excluded data taken during this time interval 
from our analysis.  According to recent {\Chandra} results \cite{Baganoff2003},
Sgr~A$^*$ is the brightest individual source in the central 10~arcsec region,
however, its contribution to the total emission from the region is only about 
10~per cent.  Since the spatial resolution of {\XMMN} is $\sim$5~arcsec 
(for the MOS cameras; Jansen {\etal} 2001), our present result is fully 
consistent with the {\Chandra} picture.

There is a further extended X-ray emission to the east from Sgr~A$^*$, which
is distinct in both the soft and hard bands.  When we exclude the 
20~arcsec-radius region around Sgr~A$^*$, the
centre of the diffuse emission is found to be located at (17$^{\rm
h}$~45$^{\rm m}$~44$^{\rm s}$, $-29\degr~0{\farcm}3$) in
the J2000 coordinates.   The e-folding radius is 28~arcsec in the core in 
the 2--10 keV band, although the enhancement in the surface brightness 
(with respect to the surrounding region) has a full extent closer to
$\sim$200~arcsec across. This diffuse feature is clearly elongated along an 
axis roughly parallel to the Galactic plane. 

As shown in Fig.~\ref{fig:img}, the X-ray emitting region is mostly confined
within the radio shell of {\SgrAEast}, in full agreement with earlier 
{\Chandra} observations \cite{Maeda2002}.  On the basis of this strong 
correlation between the extended X-ray emission and the radio shell structure, 
hereafter we refer to this bright diffuse X-ray source as {\SgrAEast}.

Since the X-ray spectrum of {\SgrAEast} shows distinct emission lines
(see Section~\ref{sec:spec}), we have made line intensity images at
6.7-keV and 2.4-keV; the former
represents the K$\alpha$ line from helium-like iron (Fe) whereas the latter 
is from helium-like sulfur (S).  We made three images corresponding to energy
bands  2.2~keV--2.6~keV (2.4-keV-band), 4.0~keV--6.0~keV
(continuum), and 6.55~keV--6.85~keV (6.7-keV-band).  From the measured
continuum image, we estimated continuum images appropriate to
the 6.7-keV and 2.4-keV bands, taking into account
the detector energy response and the continuum shape averaged
over the whole region. These narrow-band continuum images
were then subtracted to reasonable approximations to pure line
images at 6.7-keV and 2.4-keV (Fig.~\ref{fig:img}c,~d).  The 6.7-keV line is
clearly more concentrated in the core of {\SgrAEast} than the continuum
(Fig.~\ref{fig:img}c).  This implies that the core of {\SgrAEast} is more
abundant in iron, or possibly higher in temperature, or a
combination of these two effects.  In contrast, the 2.4-keV line peak is
located on Sgr~A$^*$ (Fig.~\ref{fig:img}d).
These effects are investigated quantitatively via the spatially-resolved 
spectral analysis described in the later sections
(Section~\ref{sec:separate-reg-fit}).

We note that three further bright sources  are visible in
Fig.~\ref{fig:img}(a, b).  The hard source {\XMMfil} appears to be 
a non-thermal X-ray filament (see Sakano {\etal} 2003a for details).  
The two soft sources,
{\XMMJsoft} and RX J1745.7$-$2904, have previously been detected by {\ROSAT}
(Predehl \& Tr\"{u}mper 1994; Sidoli, Belloni \& Mereghetti 2001).
{\XMMJsoft} is presumably identical with a {\Chandra} source CXOGC
J174545.2$-$285828 (Muno {\etal} 2003).
We identify RX J1745.7$-$2904 as a foreground star,
GSC 06840$-$00590\footnote{http://www-gsss.stsci.edu/gsc/gsc.html}.

\subsection[]{Spectrum \label{sec:spec}}

\subsubsection[]{Spectrum accumulation and background subtraction}

As illustrated in Fig.~\ref{fig:img}(a,b), we accumulated the source spectrum 
from a circular cell of 100~arcsec-radius but excluding the regions
within 24 and 16~arcsec-radii of Sgr~A$^*$ and {\XMMJsoft}, respectively.
The choice of an appropriate background field is not a trivial task due to the
clumpy nature of hot plasma, which pervades the whole region. Variations 
in absorption also add to the complexity of the situation.
Here, we chose a nearby background region at nearly the same galactic latitude 
as the source region (see Fig.~\ref{fig:img}), so as to minimize 
systematic effects. In fact, the X-ray hardness does not vary strongly
along this axis (Sakano {\etal} 2003b), and the difference of the X-ray
absorption between the source and background regions is also minimized.
Since {\SgrAEast} is a very bright source, the spectrum is not strongly
affected by selection of the background, except for the outer region
defined in Section~\ref{sec:separate-reg-fit}, as long as we choose the background
from a relatively nearby region.  For example, the use of
a different background dataset gave rise to an overall flux increased in
{\SgrAEast} of 5~per cent, whereas the derived temperature changed by
$\sim$10~per cent or less. Although the best-fitting values change slightly
depending on the background selection, the spectral structure never
changes. We conclude that the spectral results presented below which are
derived on the basis of the background field identified in 
Fig.~\ref{fig:img}(a) are reasonably robust.

Fig.~\ref{fig:pn-fit} shows the resultant pn spectra.  Several
emission lines can be seen, implying that a significant
fraction of the spectrum originates from hot thermal plasma.

\subsubsection[]{Fitting with a phenomenological model \label{sec:pn-fit}}

Using the {\sc xspec} package we initially experimented with the
simultaneous fitting of the MOS1 and MOS2 spectra with a ``phenomenological 
model'' consisting of a multi-temperature continuum plus many Gaussian lines. 
The continuum comprised three thermal bremsstrahlung components with 
temperatures initially set at 0.5, 1.0 and 4 keV, each absorbed by a 
separate column density component. However, in the event it proved necessary
to reduce the number of free parameters describing the continuum; this
was achieved by fixing the temperature and column for the softest component at
0.5 keV and $7 \times 10^{22} \rm~cm^{-2}$ respectively and absorbing
the two higher temperature components with the same column density. 
With continuum so defined, we were able to determine the equivalent widths 
and centre energies for the set of lines, although for most of the weak lines
we fixed the centre energies at the theoretical values expected for a 
thin-thermal plasma emission \cite{Mewe1985}.  In this  fitting process, 
in addition to  K$\alpha$ lines from highly ionized atoms, we also
included some K$\beta$ lines, but in this case fixed the line energy 
and the ratio of the normalisation between K$\alpha$ and K$\beta$ lines at the values predicted by thin-thermal plasma models 
\cite{Mewe1985}. Similarly the fluorescent lines from neutral iron
(K$\alpha$, K$\beta$) were fitted at fixed line energies.
As for the line width, we only allowed that of the 6.7-keV line, which
has by far the best statistics, to be free. The other line widths
were fixed at the values expected for a thin-thermal
plasma \cite{Mewe1985}, taking the energy resolution of the detector
into account; {\eg}, we fixed the width of the 3.13-keV line, which is
presumably a K$\alpha$-line from helium-like argon, to be 9~eV because 
this line is actually a combination of resonance, forbidden,
intercombination, and satellite lines.

\begin{figure}
 \psfig{file=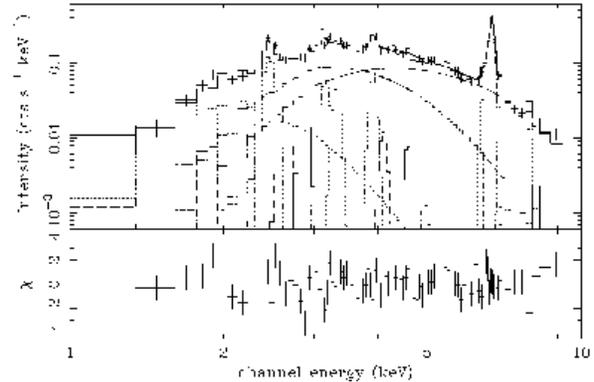,width=0.45\textwidth,angle=270,clip=}
 \caption{The measured background-subtracted EPIC/pn spectrum of Sgr~A East. 
The best-fitting ``phenomenological model'' corresponds to a multi-temperature
bremsstrahlung continuum plus multiple Gaussian lines.  
The bottom panel illustrates the spectral fitting residuals.
  \label{fig:pn-fit}}
\end{figure}

With the best-fit to the MOS spectra determined, we then fitted the pn 
spectrum with the same model.  The MOS and pn results were found to be 
consistent with each other except for slight differences in the normalisation
and energy scale (the latter plausibly attributable to a detector gain error).
The pn and MOS spectra were aligned when we applied a gain correction
of $\sim$0.1~per cent to the former. Since the calibration uncertainty 
is $\sim$0.3~per cent (Kirsch 2002), this difference is within the probable 
uncertainty.  In the analysis described below the pn gain was fixed at 1.001.

Table~\ref{tbl:fit-eachline} summarises the best-fitting results for the 
separate MOS and pn fits including
probable line identifications, while Fig.~\ref{fig:pn-fit} displays the pn
spectrum with the best-fitting model.  The quoted uncertainties in
Table~\ref{tbl:fit-eachline} and henceforth are at the 90~per cent confidence
level for one interesting parameter, unless otherwise mentioned.
In particular, the most distinct
four lines are identified with K$\alpha$ lines from He-like sulfur,
argon (Ar), calcium (Ca), and iron.

\begin{table*}
 \centering
 \begin{minipage}{140mm}
  \caption[]{Best-fitting parameter values for the phenomenological model
discussed in the text \label{tbl:fit-eachline}}
  \begin{tabular}{cccccccl}
   \hline
   Param.\footnote{The derived line energy and normalisation in units of
   keV and $10^{-4}$ photons cm$^{-2}$~s$^{-1}$, respectively.  For most
   of the Gaussian lines, the centre energy is fixed to the
   theoretically expected value, which is indicated in the the last
   column, in the fitting.}
  & \multicolumn{2}{c}{Best-fit$^a$} & \multicolumn{2}{c}{Best-fit$^a$} & \multicolumn{2}{c}{Best-fit$^a$} &
   \multicolumn{1}{c}{Identification\footnote{The value in the
   parentheses is the theoretical line-centre energy (in keV),
   assuming a temperature of 1~keV, 2.2~keV, and 4~keV for He-like lines 
   below calcium, for H-like lines below calcium, or for other lines (nickel,
   iron and calcium lines), respectively.  Most of the lines are
   combinations of unresolved narrow lines.  Thus, we define the
   centre energy as the weighted mean for these unresolved lines.}}\\
        & \multicolumn{2}{c}{MOS1+2} & \multicolumn{2}{c}{pn} & \multicolumn{2}{c}{pn+MOS1+MOS2}\\
   \hline
   Norm.(1)   & 44.6  & (0.0--79.0)  & 43.9 & (7.6--94.0)  & 15.3 & (0.0--21.5)  & Si XIII K$\alpha$ (1.86)\\ 
   \hline
   Norm.(2)   & 31.2  & (15.2--47.2) & 29.4 & (16.6--50.4) & 22.3 & (10.1--62.1) & Si XIV  K$\alpha$ (2.01)\\
   \hline
   Centre(3)  &  2.46 & (2.44--2.46) & 2.44 & (2.44--2.45) &  2.44 & (fixed) &  S XV K$\alpha$ (2.44)\\
   Norm.(3)   & 21.7  & (20.1--24.5) & 15.6 & (11.8--17.1) &  17.5 & (13.1--23.7) & Si XIV K$\beta$ (2.38)\\
   \hline
   Norm.(4)   &  2.48 & (1.12--3.18) & 1.74 & (0.61--2.85) & 1.69 & (0.79--2.91) & S XVI K$\alpha$ (2.62)\\
   \hline
   Centre(5)  &  3.11 & (3.09--3.12) &  3.15 & (3.10--3.15) &  3.13 & (fixed) & Ar XVII K$\alpha$ (3.13)\\
   Norm.(5)   &  1.57 & (1.30--2.05) &  1.18 & (0.77--1.57) &  1.36 & (1.03--1.75) & S XVI K$\beta$ (3.11)\\
   \hline
   Norm.(6)   &  0.70 & (0.46--0.95) &  0.66 & (0.31--0.91) &  0.67 & (0.46--0.89) & Ar XVIII K$\alpha$ (3.32)\\
   \hline
   Centre(7)  &  3.89 & (3.87--3.91) &  3.91 & (3.89--3.94) &  3.89 & (fixed) & Ca XIX K$\alpha$ (3.89)\\
   Norm.(7)   &  0.84 & (0.69--0.98) &  0.64 & (0.48--0.82) &  0.78 & (0.65--0.89)\\
   \hline
   Norm.(8)  &  0.06 & (0.00--0.17) &  0.09 & (0.00--0.22) &  0.08 & (0.00--0.17) & Ca XX K$\alpha$ (4.10)\\
   \hline
   Norm.(9)  &  0.16 & (0.06--0.22) &  0.17 & (0.10--0.22) &  0.16 & (0.11--0.22) & Fe I K$\alpha$ (6.40)\\
   \hline
   Centre(10) &  6.67 & (6.66--6.67) &  6.68 & (6.67--6.68) &  6.68 & (fixed) & Fe XXV K$\alpha$ (6.68)\\
   $\sigma$(eV)(10) & 37 & (15--45)  & 44    & (34--54)     & 40    & (31--48) & (17~eV is expected)\\
   Norm.(10)  &  2.11 & (1.99--2.25) &  1.79 & (1.70--1.93) &  2.07 & (1.97--2.18)& \\
   \hline
   Norm.(11)  &  0.16 & (0.09--0.25) &  0.13 & (0.06--0.19) &  0.15 & (0.09--0.20) & Fe XXVI K$\alpha$ (6.96)\\
   \hline
   Norm.(12)  &  0.05 & (0.00--0.15) &  0.00 & (0.00--0.04) & 0.00 & (0.00--0.05) & Ni XXVII K$\alpha$ (7.76)\\
   \hline
   Norm.(13)  &  0.06 & (0.00--0.16) &  0.00 & (0.00--0.05) & 0.00 & (0.00--0.07) & Ni XXVIII K$\alpha$ (8.03)\\
   \hline
   \hline
   $N_{\rm H}$ & 15.2 & (14.9--15.4) & 15.2 & (14.8--15.5) & 14.7 & (13.5--15.7) & ($10^{22}${\NHUNIT})\\
   $kT_e$      & 3.97 & (3.75--4.13) & 4.17 & (3.96--4.39) & 4.07 & (fixed) & (keV)\\
   Norm.\footnote{The unit of $10^{-12} \int n_{\rm e} n_{\rm H} dV / (4\pi D^2)$, where $n_{\rm e}$ and $n_{\rm H}$ are the electron and proton number densities (cm$^{-3}$) and $D$ is the distance of the source (cm).} 
                 & 8.48 & (8.01--8.74) & 7.35 & (6.97--7.65) & 8.52 & (8.09--8.89) & (see below$^c$)\\
   $kT_e$      & 1.01 & (0.98--1.04) & 1.00 & (0.97--1.03) & 1.01 & (fixed) & (keV)\\
   Norm.$^c$   & 106  & (100--113)   & 93.3 & (85.9--101)  & 94.8 & (74.6--118)  & (see below$^c$)\\
   \hline
   $N_{\rm H}$ & 7.00 & (fixed) & 7.00 & (fixed) & 7.00 & (fixed) & ($10^{22}${\NHUNIT})\\
   $kT_e$      & 0.50 & (fixed) & 0.50 & (fixed) & 0.50 & (fixed) & (keV)\\
   Norm.$^c$   & 58.9 & (50.9--72.1) & 61.6 & (54.4--77.5) & 70.8 & (50.1--88.9) & (see below$^c$)\\
   \hline
   Factor      &  ---  &  &  ---  &  &  0.87 & (0.85--0.89) & (pn/MOS)\\
   \hline
   $\chi^2$/dof & \multicolumn{2}{l}{234.0/204} & \multicolumn{2}{l}{123.34/81} & \multicolumn{2}{l}{389.27/314}\\
   {\FX}(pn)\footnote{Observed X-ray flux in the 2--10 keV band.} & --- & & 1.10 & & 1.09 & \multicolumn{2}{r}{($10^{-11}${\FLUXUNIT})}\\
   {\FX}(MOS)$^d$                               & 1.25 & & --- & & 1.25 & \multicolumn{2}{r}{($10^{-11}${\FLUXUNIT})}\\
   \hline
\end{tabular}
\medskip

Adding to these lines, we also included, in the fitting, the
  K$\beta$-lines of Si XIII, S XV, Fe I, Fe XXV, and Fe XXVI at the
  respective energies of 2.18, 2.88, 7.06, 7.90, and 8.21 keV, fixing
  the centre energy and the ratio of the normalisation to the
  corresponding K$\alpha$ line at the theoretical value.
\end{minipage}
\end{table*}

Table~\ref{tbl:fit-eachline} also gives the results of a simultaneous fit to
all three detectors (but with all the line energies 
fixed as indicated). The global normalisation factor between the pn and
MOS spectra was found to be 0.87 (which may be accounted for at least in part
by photons in the spectrum-accumulating region in the pn falling on a
chip gap).  We apply an appropriate correction for ``the lost pn photons'' 
in the analysis which follows.

\begin{figure}
 \psfig{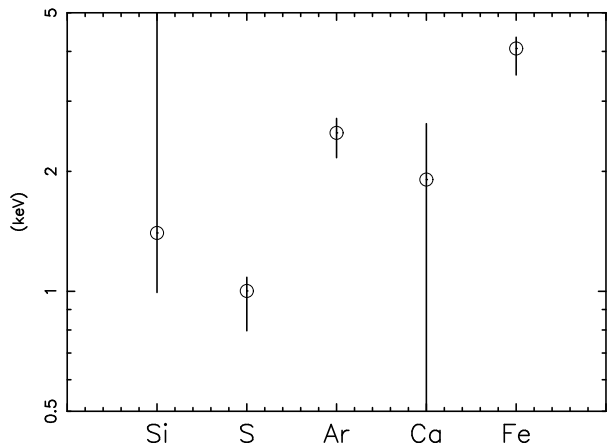}
 \caption{The ionization temperature (in keV) for five different elements 
 derived from the intensity ratio of the helium-like and hydrogen-like K-lines,
 assuming the lines originate in a thin thermal plasma in
 ionization equilibrium (Mewe {\etal} 1985).  \label{fig:line-kt}}
\end{figure}

Next we estimated the ionization temperature of the plasma from the ratio 
of He-like and 
H-like K-lines for each atomic species, using the theoretical results of
Mewe {\etal} \shortcite{Mewe1985}. 
Fig.~\ref{fig:line-kt} summarises the derived temperatures.  The plasma
temperature is found to vary significantly from atom to atom; for
example, the ionization temperatures of sulfur and iron are 1~keV and
4~keV, respectively.  This implies that the spectrum consists of
multiple temperature components.  In fact, when we tried to apply a
single-temperature thermal model to the spectrum, it is clearly
rejected.  We note that at the energy of the silicon (Si) line the continuum is 
strongly affected by absorption, hence the estimate of the ionization 
temperature of silicon 
may be subject to some additional systematic uncertainty.

\subsubsection[]{Fitting with a thermal plasma model \label{sec:simul-fit}}

\begin{figure}
  \psfig{file=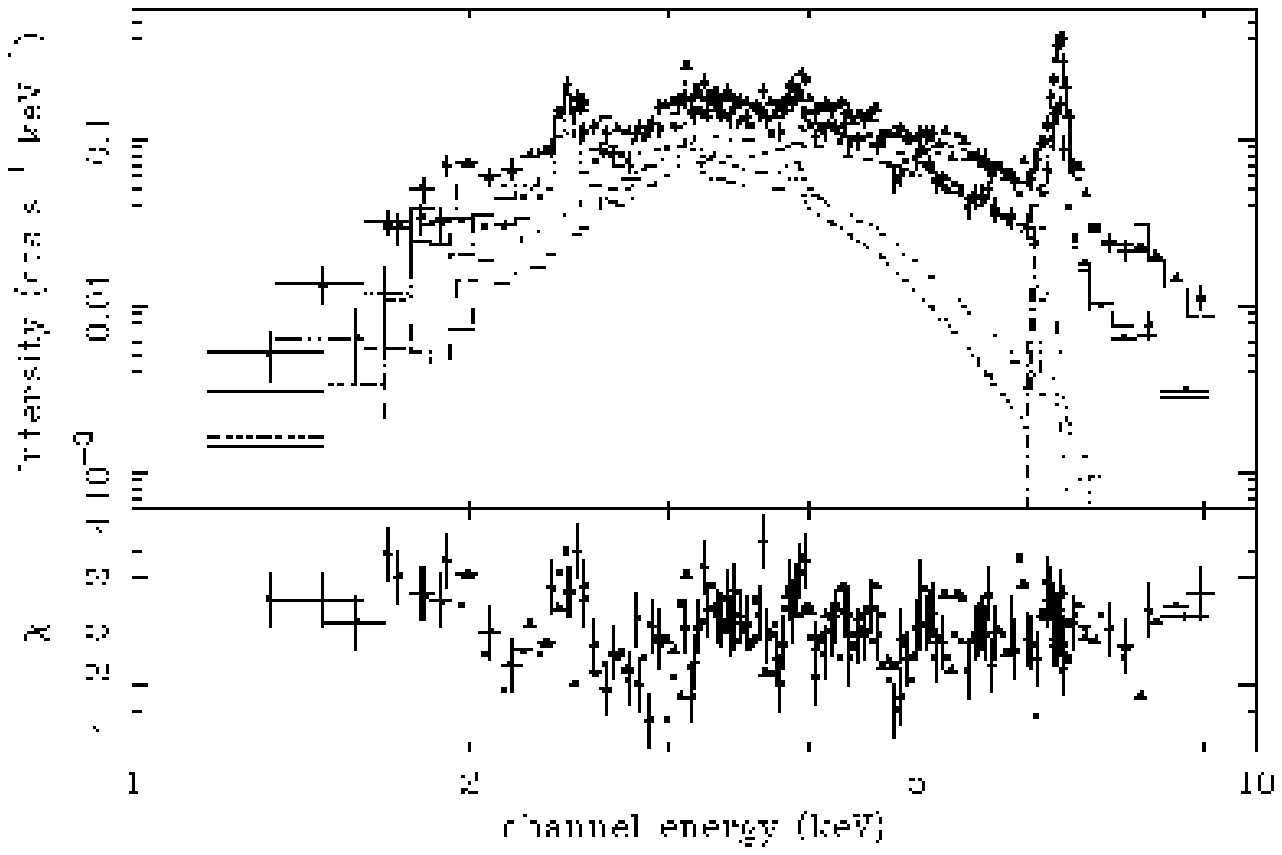,width=0.45\textwidth,angle=270,clip=}
  \psfig{file=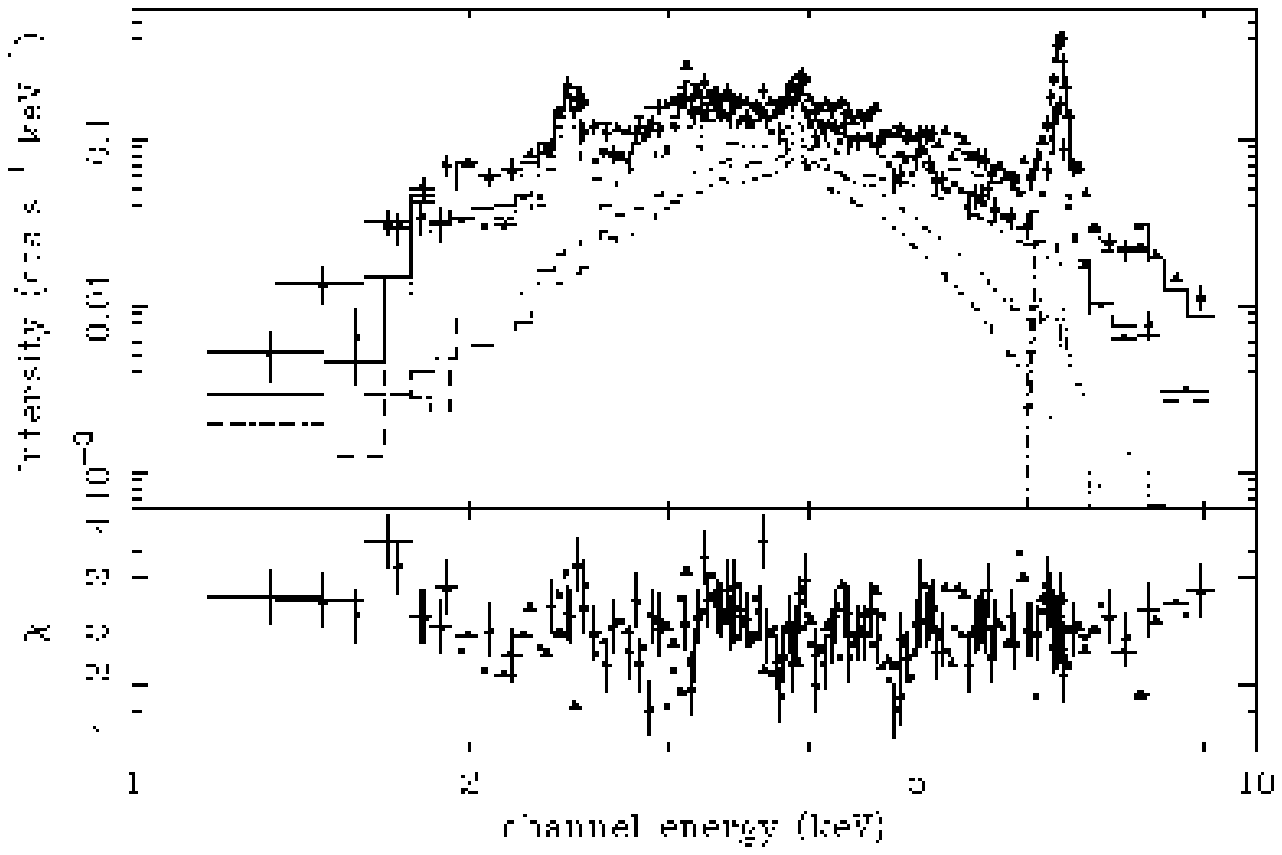,width=0.45\textwidth,angle=270,clip=}~
 \caption{The spectral fitting of Sgr A East (whole source). Measured count 
 rate spectra are plotted for the pn (filled triangle) and MOS1 (filled
 circle).  For simplicity the MOS2 data are not plotted although they are used in the fitting.
  {\it Upper Panel}: A two-component thermal plasma model with
 solar abundances assumed. {\it Lower Panel}: A two-component 
 thermal plasma model
 with the metal abundances of Si, S, Ar, Ca, and Fe allowed to vary.
 \label{fig:simul-fit}}
\end{figure}

On the basis of the results presented above, we have applied a two-temperature 
thin-thermal plasma model ({\sc mekal}) both subject to the same absorption 
component to the EPIC spectra observed from {\SgrAEast}.  We initially adopt 
the solar abundance ratio \cite{Anders1989} and, for the absorption,  assume 
the cross sections tabulated by Morrison \& McCammon (1983). The ratio of the 
normalisation between pn and MOS detector was fixed at the value obtained 
previously. Fig.~\ref{fig:simul-fit} (upper panel) shows the pn, and
MOS1 and 2 spectra with the best-fitting model, whereas the third column 
in Table~\ref{tbl:fit-kt} gives the fitting results. Notice that the residuals
to the fit are most prominent below 4~keV (Fig.~\ref{fig:simul-fit}).

As the next step in the fitting process, we allowed the relative abundances of 
silicon, sulfur, argon, calcium, and iron in the plasma to be free (but tied 
across the two different temperature components).  The abundance of
nickel was linked to that of iron.  The result was found to give a
reasonably good fit (see 
the fourth column in Table~\ref{tbl:fit-kt} and the lower panel in 
Fig.~\ref{fig:simul-fit}).
In particular the model reproduces the data above 2~keV very well.

The origin of the residuals apparent below 2~keV is not clear.
The current best-fitting model gives a silicon abundance of $\sim$5
solar, which is 2.5 times higher than that of sulfur and, hence,  may not be
realistic.  On the other hand the addition of a third (lower) temperature 
component to the spectral model does not give a significant improvement in 
the fit for reasonable ranges of temperature.  The description of the
the absorption in terms of partial covering does improved the fit
($\chi^2$/dof=359.3/320) significantly but the model parameters
are rather poorly constrained and, in any event, the 
applicability of such a model is questionable. 

In summary, the spectral results for the whole of {\SgrAEast} are:
(1) the temperatures derived for the two plasma components are
$\sim$1~keV and $\sim$4~keV, which are fully consistent with that
estimated from the line-ratios (see 
Fig.~\ref{fig:line-kt}); (2) the lower-temperature component has a much
higher emission measure than the higher-temperature one;
(3) the derived metal abundances are, on average, slightly
higher than solar values with calcium showing the highest
apparent abundance ($\sim$ 2.5 solar).

\begin{table*}
\hspace*{-26mm}
 \begin{minipage}{150mm}
  \caption{Spectral fitting results for  {\SgrAEast} and its sub-regions \label{tbl:fit-kt}}
 \begin{tabular}[tb]{cccccccc}
 \hline
  Params.
          & Unit & \multicolumn{5}{c}{--- Data ---}\\
          &      & Whole & Whole & $r<28''$ & $r=28''$--$60''$ & $r>60''$\\
 \hline
\multicolumn{2}{l}{Norm(pn/MOS)}
                     & 0.87 (fixed)      & 0.87 (fixed)      & 1.03 (1.00--1.08) & 0.78 (0.75--0.81) & 0.86 (0.82--0.90)\\
{\NH} &({\NHUNIT})& 11.8 (11.3--12.3)    & 13.5 (12.8--14.2) & 15.7 (15.3--16.0) & 14.4 (14.1--14.7) & 11.1 (10.8--11.3)\\
$kT_e$(1) & (keV)& 3.81 (3.52--4.06)     & 4.23 (3.82--4.71) & 3.05 (2.85--3.17) &  5.5 (5.1--6.1)   & 4.41 (4.20--4.89)\\
Norm(1)\footnote{Normalisation in unit of $10^{-12} \int n_{\rm e} n_{\rm H} dV / (4\pi D^2)$, where $n_{\rm e}$ and $n_{\rm H}$ are the electron and proton number densities (cm$^{-3}$) and $D$ is the distance of the source (cm).}
          &      & 2.29 (2.15--2.59)     & 1.66 (1.33--1.93) & 0.41 (0.36--0.43) & 0.37 (0.35--0.39) & 0.80 (0.74--0.84)\\
$kT_e$(2) & (keV)& 0.88 (0.81--0.95)     & 0.94 (0.85--1.00) & 0.91 (0.88--0.95) & 1.03 (0.98--1.08) & 0.92 (0.84--0.97)\\
Norm(2)$^a$&     & 12.1 (9.96--17.3)     & 11.6 (9.5--12.7)  &  2.7 (2.5--2.9)   &  4.2 (4.0--4.4)   &  3.8 (3.2--4.0)\\
$Z_{\rm Si}$\footnote{Abundances relative to solar for different atoms.}
                & & 1 (fixed) & 5.1  (3.5--7.3)   & 8.9 (7.5--11.9) & 5.6 (4.4--6.9)   & 2.7  (2.2--3.3)\\
${Z_{\rm  S}}^b$& & 1 (fixed) & 1.94 (1.56--2.47) & 2.7 (2.3--3.3)  & 2.0 (1.7--2.3)   & 1.6  (1.3--1.8)\\
${Z_{\rm Ar}}^b$& & 1 (fixed) & 1.23 (0.84--1.67) & 1.8 (1.1--2.5)  & 1.4 (0.89--1.8)  & 1.0  (0.58--1.5)\\
${Z_{\rm Ca}}^b$& & 1 (fixed) & 2.54 (2.04--3.07) & 2.5 (1.7--3.3)  & 2.6 (1.9--3.2)   & 2.9  (2.1--3.5)\\
${Z_{\rm Fe,Ni}}^b$&&1 (fixed)& 1.33 (1.18--1.51) & 3.8 (3.5--4.0)  & 1.50 (1.37--1.65)& 0.47 (0.39--0.54)\\
Fe-K$\alpha$\footnote{Intensity of the fluorescent K$\alpha$ line from neutral iron at 6.4~keV.}
 & ($10^{-5}$ph s$^{-1}$ cm$^{-2}$) & 1.8 (1.0--2.1) & 1.4 (1.1--2.2) & 0.3 (0.1--0.5) & 0.3 (0.1--0.7) & 1.0 (0.6--1.3)\\
 \hline
$\chi^2$/dof &   & 459.9/326 & 374.5/321 & 203.5/182 & 241.8/232 & 207.7/198\\
$F_{\rm X}$\footnote{Observed flux in the 2--10 keV band.} & ($10^{-12}$erg s$^{-1}$
  cm$^{-2}$) & 12.6 & 12.6 & 3.2 & 4.2 & 5.0\\
 \hline
 \end{tabular}

\end{minipage}
\end{table*}

\subsubsection[]{Spectral variation within Sgr A East 
\label{sec:separate-reg-fit}}

Fig.~\ref{fig:img} (c) shows that the He-like iron K-line emission shows
a greater concentration in the core region of {\SgrAEast} than is apparent in
the 2--10 keV image (Fig.~\ref{fig:img}a). To investigate 
spectral variations within {\SgrAEast}, we have extracted spectra from 
three sub-regions of the source (see Fig.~\ref{fig:img}c), namely the core (a circle region
with $ r < 28$~arcsec centred on the peak of the 6.7-keV line), the central annulus 
(with $r = $28--60~arcsec again centred on 6.7-keV line peak) and the outer region
({\ie} the whole source region except for the core and central 
annulus).  We fitted the spectra of these sub-regions with the same
model described in the previous sub-section, allowing the relative
normalisations for pn and MOS detectors to vary.

\begin{figure}
  \psfig{file=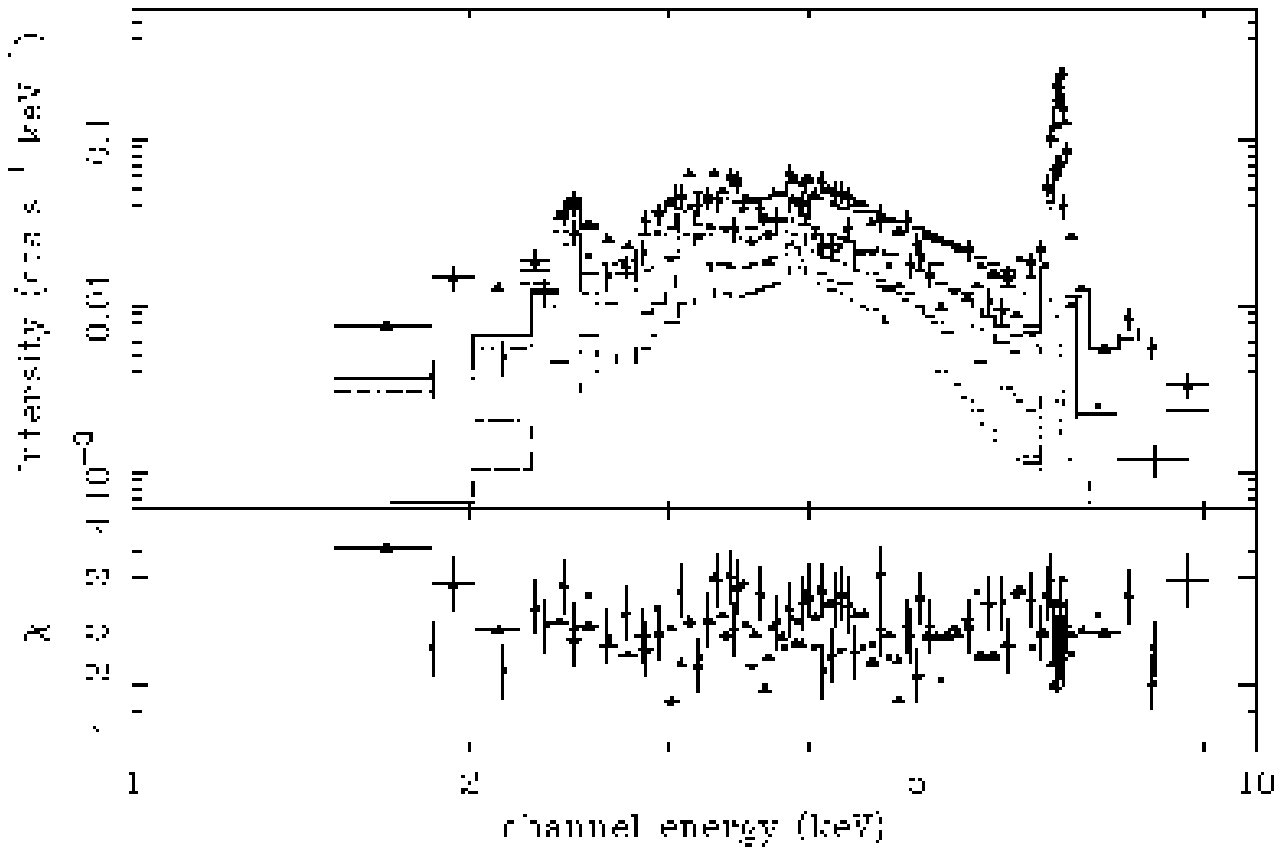,width=0.45\textwidth,angle=270,clip=}
  \psfig{file=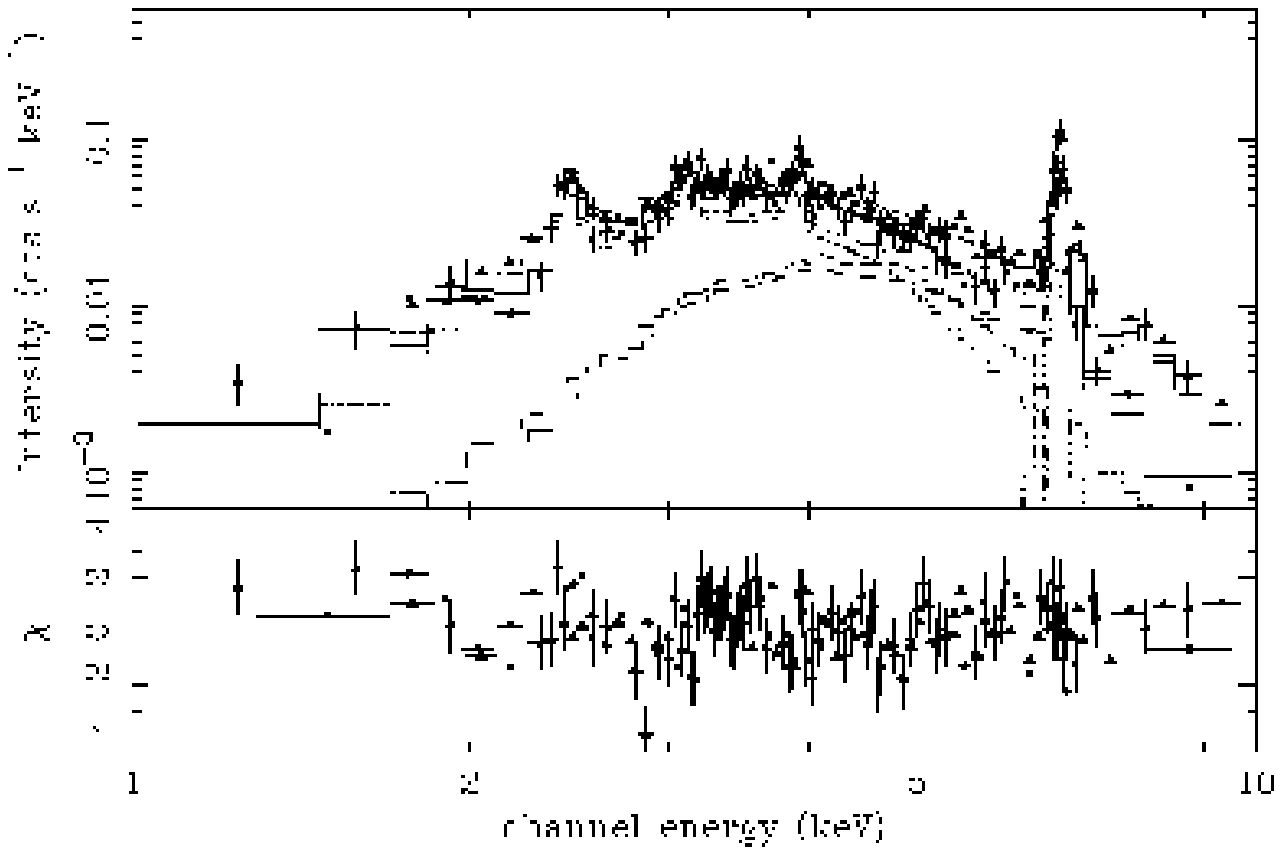,width=0.45\textwidth,angle=270,clip=}
  \psfig{file=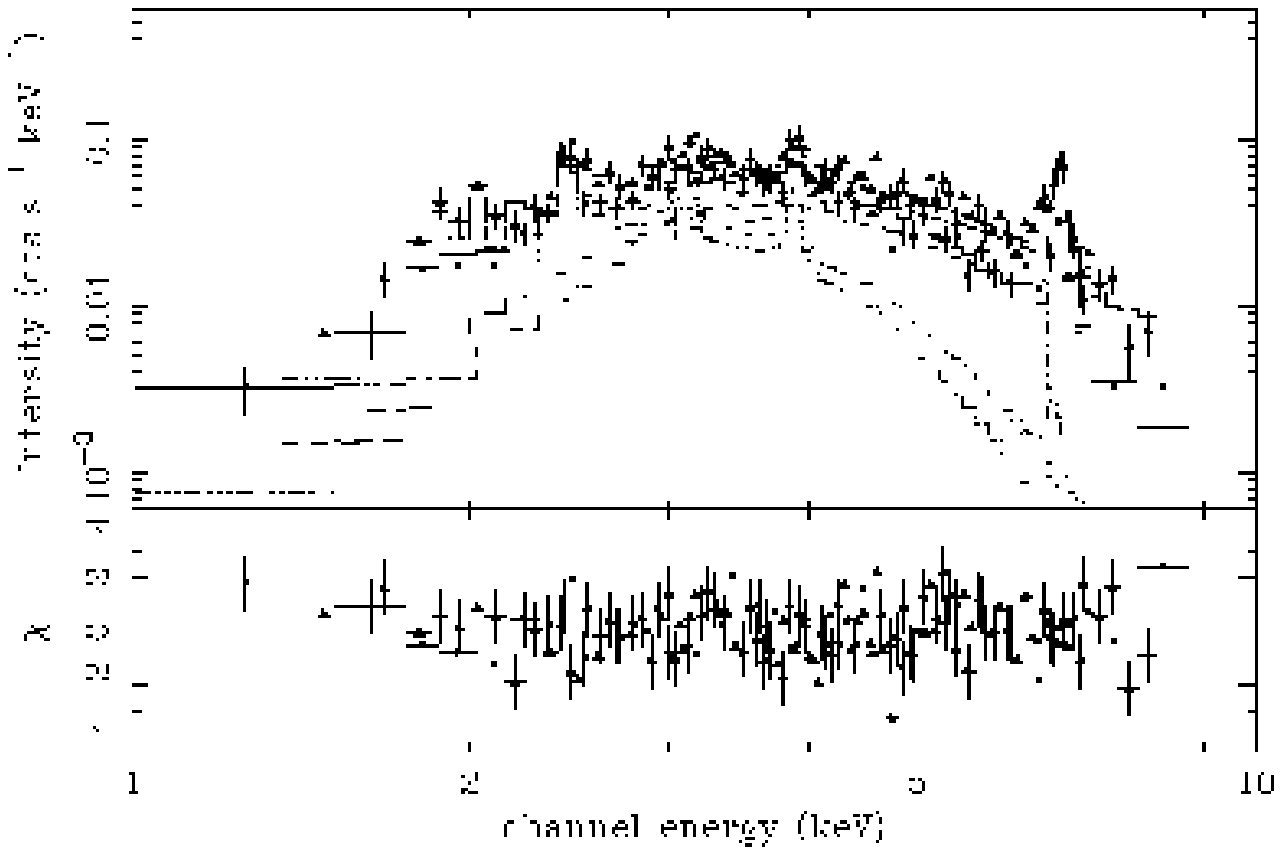,width=0.45\textwidth,angle=270,clip=}~
 \caption{The fitting of the two-temperature thin thermal plasma 
 model with patchy absorption to the spectra measured from three separate 
 regions of Sgr A East.  The notation is the same as that of Fig.~\ref{fig:simul-fit}.
 {\it Top Panel:} the core region. {\it Middle Panel:}
 the central annulus. {\it Bottom Panel:} the outer region of the source.
 \label{fig:separate-reg-fit}}
\end{figure}

Fig.~\ref{fig:separate-reg-fit} shows the measured spectra together with
the best-fitting models, and the last three columns of
Table~\ref{tbl:fit-kt} lists the corresponding best-fitting parameter
values.  Consistent with the image analysis, we find that the iron abundance 
is highest in the core region ($Z_{\rm Fe} \approx 3.3$), dropping to about half 
this value in the central annulus to only $Z_{\rm Fe} \approx 0.5$ in the outer 
region. In contrast the abundances of the other metals (S, Ar, and Ca) do not show large
spatial variation across {\SgrAEast}.  The derived plasma temperatures are 
slightly lower in the core region but the ratio of the emission
measures of the lower- and higher-temperature components does not
change.  The 6.4-keV iron fluorescent line is highly significant particularly in the outer 
region of {\SgrAEast}, where the equivalent width is $\sim$85~eV.

\subsubsection[]{Plasma parameters \label{sec:plasma}}

The line-of-sight column density measured for {\SgrAEast} is 
$\sim 1.5 \times 10^{23}${\NHUNIT}.  The column density measured
for X-ray point sources in the {\GC} region is found, in general, to 
be inversely correlated with the angular distance from the Galactic plane 
(Sakano {\etal} 1999; Sakano 2000) and for the position of {\SgrAEast}, 
the measured column would typically be in excess $10^{23}$ {\NHUNIT}.  In
fact, the foreground column density is probably only $3\times
10^{22}$ {\NHUNIT} (Sakano 2000),  hence based on column density
arguments alone one would place {\SgrAEast} truly in the {\GC} region. 
We therefore assume the distance of {\SgrAEast} to be that of the {\GC}, 
namely 8.0~kpc \cite{Reid1993}.  Many other observational results, mostly from
radio observations, also support this idea ({\eg} Yusef-Zadeh {\etal} 2000).

The apparent extent of the core of {\SgrAEast} in the hard X-ray band is 
28~arcsec in radius (see Section~\ref{sec:image}).  Hence, we calculate the total 
plasma volume $V$ to be 1.6$\times 10^{56}$~cm$^3$, assuming a spherical 
shape. Hereafter we use the best-fitting parameters for the core region 
($r<$28~arcsec in Table~\ref{tbl:fit-kt}), unless otherwise stated.
We estimate the emission measure (EM) of the lower- and higher-temperature 
components to be
\begin{eqnarray}
 {\rm EM}_{\rm L} \equiv\, n_{\rm e,L}^2 \eta_{\rm L} V & \approx &\, 20\,\times 10^{57}  \eta_{\rm L}  \ \ \ ({\rm cm}^{-3}) \label{eq:EM_L}\\
 {\rm EM}_{\rm H} \equiv   n_{\rm e,H}^2 \eta_{\rm H} V & \approx &  3.1\times 10^{57} \eta_{\rm H}  \ \ \ ({\rm cm}^{-3}),
\end{eqnarray}
respectively, where $n_{\rm e}$ is an electron density, $\eta$ is a
filling factor and $V$ is the volume occupied by both plasma components 
(as estimated above). The indices L and H refer to the lower- and 
higher-temperature components, respectively.

The filling factors of the two components are unknown.  Plasmas with two
apparently different temperatures, however, are not likely
to co-exist in the same volume, so the sum of the two filling factors
for both the plasmas should be $\le 1.0$.  If we assume a
pressure balance, then
\begin{eqnarray}
 n_{\rm e,L} T_{\rm L} = n_{\rm e,H} T_{\rm H},
\end{eqnarray}
where $T$ is the plasma temperature. In the present case, 
$T_{\rm L} = 0.91$ keV and $T_{\rm H} = 3.0$ keV (Table~\ref{tbl:fit-kt}).

Then, introducing the total filling factor $\eta_{\rm tot}$,
\begin{equation}
\eta_{\rm tot} \equiv \eta_{\rm L} + \eta_{\rm H}, \label{eq:eta_tot}
\end{equation}
and solving the equations (\ref{eq:EM_L}--\ref{eq:eta_tot}), we find
\begin{eqnarray}
\eta_{\rm L} & = & \frac{\eta_{\rm tot}}{1+\frac{{\rm EM}_{\rm H}}{{\rm EM}_{\rm L}}\left(\frac{T_{\rm H}}{T_{\rm L}}\right)^2}\ \ \ \approx 0.37\ \eta_{\rm tot}\\
\eta_{\rm H} & = & \frac{\eta_{\rm tot}}{1+\frac{{\rm EM}_{\rm L}}{{\rm EM}_{\rm H}}\left(\frac{T_{\rm L}}{T_{\rm H}}\right)^2}\ \ \ \approx 0.63\ \eta_{\rm tot}\\
n_{\rm e,L} & = & \sqrt{\frac{{\rm EM}_{\rm L} + {\rm EM}_{\rm H}\left(\frac{T_{\rm H}}{T_{\rm L}}\right)^2}{\eta_{\rm tot} V}}\ \ \ \approx \frac{19}{\sqrt{\eta_{\rm tot}}}\ \  ({\rm cm}^{-3})\\
n_{\rm e,H} & = & \sqrt{\frac{{\rm EM}_{\rm H} + {\rm EM}_{\rm L}\left(\frac{T_{\rm L}}{T_{\rm H}}\right)^2}{\eta_{\rm tot} V}}\ \ \ \approx \frac{5.6}{\sqrt{\eta_{\rm tot}}}\ \  ({\rm cm}^{-3}).
\end{eqnarray}

Using the above values, the pressure ($P$), energy of the plasmas
($E_{\rm L}$, $E_{\rm H}$), and mass ($M_{\rm L}$, $M_{\rm H}$) are calculated as follows,
\begin{eqnarray}
 P & = & n_{\rm e,L} T_{\rm L}\ \ \ (= n_{\rm e,H} T_{\rm H})\\
   & \approx & 2\times 10^8\ \sqrt{{\eta_{\rm tot}}^{-1}}\ \ ({\rm K}\ {\rm cm}^{-3})\\
 E_{\rm L} & = & 3 n_{\rm e,L} kT_{\rm L} \eta_{\rm L} V\ \ \ \approx \ \ 5\times 10^{48}\ \sqrt{\eta_{\rm tot}}\ \ ({\rm erg}) \label{eq:ene_low}\\
 E_{\rm H} & = & 3 n_{\rm e,H} kT_{\rm H} \eta_{\rm H} V\ \ \ \approx \ \ 8\times 10^{48}\ \sqrt{\eta_{\rm tot}}\ \ ({\rm erg})\\
 M_{\rm L}  & = & n_{\rm e,L} \eta_{\rm L} V\ \ \ \approx \ \ 0.9\ \sqrt{\eta_{\rm tot}}\ \ ({\rm M_{\sun}})\\
 M_{\rm H}  & = & n_{\rm e,H} \eta_{\rm H} V\ \ \ \approx \ \ 0.5\ \sqrt{\eta_{\rm tot}}\ \ ({\rm M_{\sun}}). \label{eq:mass_hgh}
\end{eqnarray}

The inferred total thermal energy $E_{\rm tot}$ and total mass in the hot 
plasma is, accordingly, $\sim 1.3\times 10^{49}$ erg and 1.4
M$_{\sun}$, respectively.  Note that we assume solar abundance
ratios in the calculation of the mass.

The above estimates related to the core region of {\SgrAEast}.  If we use 
the best-fitting parameters for the whole source (the fourth column in 
Table~\ref{tbl:fit-kt}) and assume $\eta_{\rm tot}$ to be 0.02
($\simeq (28''/100'')^3$), the derived values are not appreciably changed; 
for example, the derived total
thermal energy and total mass are $\sim 3.0\times 10^{49}$ erg and 2.4
M$_{\sun}$, respectively.  We note that these values may be larger,
depending on the filling factor $\eta_{\rm tot}$
(eq.~\ref{eq:ene_low}--\ref{eq:mass_hgh}).

\section[]{DISCUSSION}

\subsection[]{The {\SgrAEast} X-ray source}

The X-ray spectrum of {\SgrAEast} shows many emission lines from
highly ionized atoms.  The line-ratio analysis and the spectral fitting 
give a consistent view, namely that there are (at least) two
temperature components present at $kT\sim$1~keV and 4~keV. 
We explain a modest soft spectral excess in terms of patchy absorption.
Averaged over the whole source, the derived abundances for different
elements are in the range 1--3 times solar. However, there is a significant
concentration of iron in the core of the source, amounting to a factor
$\sim$4 overabundance with respect to solar norm.

Koyama {\etal} \shortcite{Koyama1996}, Sidoli \& Mereghetti
\shortcite{Sidoli_M1999}, and Maeda {\etal} \shortcite{Maeda2002} have
presented the {\ASCA}, {\SAX} and {\Chandra} results of {\SgrAEast},
respectively.  Our results are in general consistent with the published
findings but also give further constraints on the nature of the source.

On the basis of {\ASCA} data, Koyama {\etal} \shortcite{Koyama1996} claimed 
that the X-ray spectral characteristics of {\SgrAEast} are very similar to 
that of the hot plasma which extends over the whole of the Galactic Centre 
region ({\ie} over a $\pm 1$ deg scale).   Our {\XMM} results confirm
the presence of a multi-temperature hot plasma in  {\SgrAEast} which,
at least in qualitative terms, matches the spectral form seen elsewhere
in the region. However, Tanaka {\etal} \shortcite{Tanaka2000} report that the 
helium-like and hydrogen-like iron lines in the integrated spectrum of the 
whole {\GC} hot plasma are broadened by 73$\pm$14~eV, whereas we find little
evidence for broadening of these lines in the spectrum of {\SgrAEast}.

The higher spatial resolution afforded by {\Chandra} is an advantage for
Galactic Centre studies in that much fainter point sources can  be
identified and their contribution excluded in spectral accumulation.  
In fact, in the case of {\SgrAEast} the total contribution of discrete 
sources is less than 1~per cent of the diffuse emission 
(Y. Maeda; private communication), hence such sources represent a negligible
contamination of the {\XMMN} data. Using {\Chandra},
Maeda {\etal} \shortcite{Maeda2002} find that the spectrum of
{\SgrAEast} can be approximated as a thin-thermal plasma model with a
temperature of 2~keV, whereas we require temperatures of 1 and
4~keV.  We regard these results as consistent, given that the {\XMM}
spectra have much better statistics\footnote{  
We have tested whether our results are dependent on the choice
of background region. If we use the same background region
as employed by Maeda {\etal} \shortcite{Maeda2002} we still have
incontrovertible evidence for a two-temperature plasma, although the
inferred flux in the outer region drops by 50\%. The latter
is a consequence of the background region being
much smaller and closer to Sgr~A$^*$. However, the result for the core 
region is much less affected ($<$10\% change in flux).}.
Maeda {\etal} \shortcite{Maeda2002} 
also find that the iron line is significantly stronger in the inner region 
of {\SgrAEast}.  We confirmed this result and are able to track a decline
in iron abundance from the core to the outer region of the source.

 Further investigation shows
that the X-ray spectrum of the outer region of {\SgrAEast} is, in fact,
remarkably similar to that measured for the region immediately
around Sgr~A$^*$ (which we have specifically excluded in
the present analysis -- but see Sakano {\etal} 2003b).  In particular, the 
inferred iron abundances are very consistent ({\ie} at least
a factor three below that measured for the core region of {\SgrAEast}).  
Given this spectral similarity could these two structures have the same 
origin?  Koyama {\etal} \shortcite{Koyama1996} have suggested that
the high-temperature plasma which pervades the whole {\GC} region
might in fact be linked in some way to past activity in Sgr~A$^*$. However,
the only realistic mechanism for heating and ionizing the surrounding 
media would seem to be episodic outbursts of the central black hole and it is
far from clear how the observed X-ray morphology of the {\GC} region
could arise in such a scenario. In what follows,  we interpreted the whole of 
{\SgrAEast} as one coherent object.

The present {\XMMN} observations demonstrate very clearly
that {\SgrAEast} (and particularly its core region)
is not bright in the 6.4 keV iron fluorescence line, which characterises 
regions such as Sgr B2 (Murakami {\etal} 2000; Murakami, Koyama 
\& Maeda 2001b), where the density of cold molecular material is 
particularly high (see below).
Murakami {\etal} (2000, 2001a, b) argue that the 6.4-keV line is produced
as fluorescent emission on the surface of dense molecular clouds 
which are illuminated by some external source (or possibly one
or more intense embedded sources). An alternative scenario is that
the X-ray fluorescence is excited by cosmic rays ({\eg} Valinia {\etal} 2000).
Our measurements show that the 6.4-keV line in {\SgrAEast} is
significant with the overall equivalent of $\sim$30~eV.  In the outer
region the equivalent width of the 6.4-keV line (85$^{+26}_{-34}$~eV) is
significantly stronger than the inner region ($<$25~eV in the core region)
(see Section~\ref{sec:separate-reg-fit}).
This outer region is nearly coincident with the
observed dust ring/shell (Mezger {\etal} 1986, 1989), suggesting the
possibility that the dust ring is the source of the line. 
Mezger {\etal} \shortcite{Mezger1989} reported that the dust shell has
a density of 10$^4$~cm$^{-3}$ on average.  Then, for a thickness of 1~pc,
the column density of the shell is 3$\times 10^{22}${\NHUNIT}.  On the
other hand, to account for the equivalent width of (50--110)~eV in the 
highly simplified setting of a central illuminating X-ray source 
emitting a bremsstrahlung with a temperature of 4~keV and surrounded 
by an isotropic absorbing medium, a column density of $(4-9)\times
10^{22}${\NHUNIT} is required.  Clearly
these two estimates of the column are sufficiently close for the dust shell 
model to be considered a very viable explanation of the observed iron
fluorescence emission. Presumably in this scenario the illuminating source
is {\SgrAEast} itself.

\subsection[]{Is {\SgrAEast} a SNR?}


The observed radio-shell and X-ray centre-filled morphology suggest that 
{\SgrAEast} may be a so-called mixed-morphology SNR \cite{Rho_P1998}. 
Since the X-ray morphology is not shell-like, estimating
the age of the SNR from the X-ray data is not a trivial task.  
For the 1 and 4~keV plasma components, the velocities of sound 
are roughly 490 and 900~km~s$^{-1}$, respectively.  It follows
that the sound-crossing times within the radio shell (a radius of $\sim$4~pc) 
for the two temperature components are 7700 and 4200 yr, respectively,
which are at least zeroth order estimates of the SNR age.

The centre energy of each line (Table~\ref{tbl:fit-eachline}) suggests
that the plasma has already reached a state of ionization equilibrium.
The ionization parameters $n_{\rm e}t$ for both the temperature components are
calculated to be $\sim 6\times 10^{12} t_4 \eta^{-\frac{1}{2}}_{\rm
tot}$~cm$^{-3}$~s and $\sim 2\times 10^{12} t_4
\eta^{-\frac{1}{2}}_{\rm tot}$~cm$^{-3}$~s,
where $t_4$ is the time since the explosion in units of $10^4$ yr, and
$\eta_{\rm tot}$ is the total filling factor defined in
Section~\ref{sec:plasma}.  This means that $t\ga 5000$ yr is required
for the establishment of ionization equilibrium using the estimated
densities.  Thus, the larger of the two age estimates quoted above is 
consistent with the spectral constraints.
Mezger {\etal} \shortcite{Mezger1989} and Uchida {\etal}
\shortcite{Uchida1998} independently estimated the age of {\SgrAEast},
basically from the radio data, to be 7500 and 10000 yr leading to
a consistent overall picture.

{\SgrAEast} was found to comprise a two-temperatures component. Most likely,
this is only a rough approximation of a reality in which there are
many plasma blobs present with temperatures in the range 1--4 keV 
(or even lower).  Since there is little
temperature variation across {\SgrAEast} (see Sakano {\etal} 2003b),
the size of each blob is probably relatively small, $<$0.5~pc.  
Adopting this blob model, the filling factor ($\eta_{\rm tot}$) 
can be smaller than unity.

The total energy of $\sim 1.3\times 10^{49} {\eta_{\rm tot}}$ erg
derived from the core region is smaller than the nominal energy for a
SNR ($\sim 10^{51}$ erg).  Even if we take into account the full extent
of the SNR, the total energy is less than 10$^{50}$ erg.  Since
we believe that the plasma has already reached ionization equilibrium, the 
estimated energy should represent the whole observed thermal energy in the hard
X-ray band.  Part of the emission in {\SgrAEast} might have already
cooled, possibly due to localised, very high-density clouds. Alternatively
a significant portion of the energy might have been used for particle 
acceleration, which has as its signature GeV $\gamma$-rays \cite{Mayer1998}.  
In any event, there is no requirement for multiple supernova explosions
in interpreting {\SgrAEast} as a SNR.  The total energy we
derive here is consistent with the {\Chandra} measurements \cite{Maeda2002}, 
but is smaller by 3 orders of magnitude than the estimates quoted by 
Mezger {\etal} \shortcite{Mezger1989}.

The estimated mass of 1.4${\eta_{\rm tot}}^{\frac{1}{2}} {\rm M}_{\sun}$ 
in the hot plasma in the core region is consistent with the SNR hypothesis.
For an age of 8000 yr, this mass of the core region could be largely that of the ejecta or 
could be largely swept-up interstellar matter. In the latter case
the concentration of iron in the central 2~pc region of {\SgrAEast}
is problematic.  If the overabundance is the result of past supernova
activity over a timescale of say several million of years, then
the extent of the enriched region would be larger than 50~pc. However, the 
effect of Galactic sheer induced by differential Galactic rotation is also 
critical on such a timescale (see Lugten {\etal} 1986).
We note that the total X-ray emitting mass from the full extent of this
SNR could be $\sim$10~M$_{\sun}$, when we use the parameters from the
whole region for the estimates, and
this mass should be mostly swept-up mass.

A remarkable feature in the spectrum of the core region is the
over-abundance of iron, $Z_{\rm Fe}\sim 4$ solar, whereas the abundances
of the lighter elements stay at 1--3 solar.  The current fitting is
based on the emission integrated along the line of sight.  However, we
estimated the true iron abundance of the core region inside a radius of
1.1~pc to be $Z_{\rm Fe}\sim 5$ solar, once its contribution of the
annular regions, {\ie} in front of and behind the core region, are
subtracted.  On the other hand, the abundances of other lighter elements
(sulfur, argon, and calcium) show the rather uniform distribution.
  Therefore, the abundance of iron in the actual
core region of a radius of 1.1~pc relative to the lighter elements can
be 2--3 times solar.

The progenitor of type-Ia supernovae is, theoretically, more abundant in
iron than lighter elements ({\eg} Tsujimoto {\etal} 1995; Nomoto {\etal}
1997b).  On the other hand, the iron abundance in type-II supernovae is
known to be a strong function of the mass of the progenitor star; {\ie}
as the mass increases, the iron abundance reduces, with iron becoming
less abundant than lighter elements if the mass is $> 20\ {\rm M}_{\sun}$
({\eg} Thielemann, Nomoto \& Hashimoto 1996; Nomoto {\etal} 1997a).
Taking into account the uncertainty of both the observation and models,
we conclude that {\SgrAEast} is a remnant of either a type-Ia supernova
or a type-II event with the mass of the progenitor star of $\la 20 {\rm
M}_{\sun}$ in the latter case.  Our estimated mass (see
Section~\ref{sec:plasma}) is consistent with both of these scenarios
because a significant part of the X-ray emitting material can be a
swept-up mass.

Our findings of a fairly uniform temperature distribution and of a sharp
density increase towards the core of the source, as well as the indication
from radio observations of an interaction between the shell and
the nearby molecular cloud, are in good agreement with the observed
properties of other mixed-morphology SNRs \cite{Rho_P1998}, as already 
discussed by Maeda {\etal} \shortcite{Maeda2002}.  
One major difference in {\SgrAEast} from other mixed-morphology SNRs is
that the X-ray emission (of the core region) is associated mostly with the ejecta.  Another 
difference is the unusually high temperature of 4~keV.  In fact, this 
temperature is comparable with, or even higher than, the temperatures of young
`historical' SNRs such as Cas~A (Holt {\etal} 1994; Hughes {\etal} 2000),
Kepler \cite{Kinugasa1999} and Tycho (Hwang, Hughes \& Petre 1998;
Decourchelle {\etal} 2001).  Although the cause of this unusually high
temperature is unknown, the special environment in the {\GC} region may
be responsible.

\section[]{SUMMARY}

From a recent {\it XMM-Newton} observation of {\SgrAEast}, we obtain the
following results;
\begin{enumerate}
 \item The measured X-ray spectrum probably implies the presence of
 a multi-temperature thin-thermal plasma but this can be approximated as a two 
temperature system  with $kT \approx 1$~keV and 4~keV.
 \item The abundance averaged over a 100~arcsec radius ($\sim$4~pc) region
       is somewhat higher than solar for atoms from sulfur to iron.
 \item The iron abundance is high in the core region 
of the source (where $Z_{\rm Fe} \sim 4$) but less than the solar value 
in the outer region.
 \item The volume density is estimated to be 19 electrons cm$^{-3}$
       for the 1-keV plasma and  6 electrons cm$^{-3}$ for the 4~keV 
       component. The total energy and X-ray emitting mass of the core
       region are estimated to be $\sim 1.3\times 10^{49}$ erg and 1.4
       M$_{\sun}$, respectively. Here we have assumed the pressure balance 
       between the different temperature components and the total filling 
       factor of 1 within a spherical volume of radius 1.1~pc.
 \item The abundance pattern of the plasma from sulfur to iron suggests
       that the X-ray emission originated in a Type-Ia supernova
       explosion or Type-II one in a relatively low-mass progenitor star.
 \item Fluorescent 6.4 keV iron line emission is observed.
       Plausibly the illumination source is {\SgrAEast}
       itself.
\end{enumerate}

\section*{Acknowledgments}

The authors express their thanks to all the staff involved, past and present,
in the development, operation  and support of {\XMMN} mission.  We are 
grateful to Dr. R. Willingale, Dr. T. Tsujimoto and the anonymous referee for their valuable comments, and to Dr. Y. Maeda 
and Dr. S. Park for sharing their knowledge of the {\Chandra} data.
We thank Mr. R. Saxton, Dr. S. Sembay, Dr. G. Griffiths, and
Dr. I. Stewart for their help with matters relating to the calibration and 
analysis software.  M. S. acknowledges the
financial support from the Japan Society for the Promotion of Science
(JSPS) for Young Scientists.

\bsp

\label{lastpage}

\end{document}